\documentclass[aps,prb,twocolumn,groupedaddress,showpacs,reprint]{revtex4-2}

\usepackage{graphicx}
\usepackage{dcolumn}
\usepackage{bm}
\usepackage{physics}
\usepackage{adjustbox}
\usepackage{tabularx,multirow}
\usepackage{siunitx}
\usepackage{booktabs,array,ifthen}
\usepackage{threeparttable,array}
\usepackage{ragged2e}

\newcommand{\tdrule}{\specialrule{0.3pt}{0pt}{1.2pt}%
            \specialrule{0.3pt}{0pt}{\belowrulesep}}
\newcommand{\dtrule}{\specialrule{0.3pt}{0pt}{1.2pt}%
            \specialrule{0.3pt}{0pt}{\aboverulesep}}
\newcolumntype{P}[1]{>{\centering\arraybackslash}p{#1}}
\newcolumntype{M}[1]{>{\centering\arraybackslash}m{#1}}
\newcolumntype{R}[1]{>{\raggedright\arraybackslash}m{#1}}

\begin{document}


\title{Adaptively compressed exchange in LAPW}
\author{Davis Zavickis}
\author{Kristians Kacars}
\author{J\={a}nis C\={\i}murs}
\author{Andris Gulans}
\affiliation{Department of Physics, University of Latvia, Jelgavas iela 3, Riga, LV-1004 Latvia}

\date{\today}

\begin{abstract}
We present an implementation of the adaptively compressed exchange (ACE) operator in the linearized-augmented plane waves formalism.
ACE is a low-rank representation of the Fock exchange that avoids any loss of precision for the total energy. 
Our study shows that this property remains in the all-electron case, as we apply this method in non-relativistic total-energy calculations with a hybrid exchange-correlation functional PBE0.
The obtained data for light atoms and molecules are within a few $\mu$Ha of the precise multi-resolution-analysis calculations. 
Aside from ACE, another key ingredient to achieve such a high precision with Fock exchange was the use of high-energy local orbitals.
Finally, we use this implementation to calculate PBE0 gaps in solids and compare the results to other all-electron results.
\end{abstract}

\maketitle


\section{\label{sec:intro}Introduction}

Over the past decades, density-functional theory (DFT)~\cite{Hohenberg1964,Kohn1965} has become a Swiss-army tool in first-principle simulations of solids. 
The initial success was built on the local density approximation and was further solidified by the generalized-gradient approximation (GGA).  
The GGA parametrization introduced by Perdew, Burke and Ernzerhof (PBE) ~\cite{Perdew1996} has become the most widely used exchange-correlation functional at present, as it is not only used in studies focused on individual materials, but also a method of choice for high-throughput studies\cite{Jain2013,Kirklin2015,Curtarolo2012}.

Despite the success in many applications, GGA's predictive power is limited for atomization energies, bond lengths and the band gap in semiconductors and insulators. 
A common approach for improving the accuracy of computational tools is to combine GGA with the Hartree-Fock (HF) approximation expressing the exchange-correlation energy as 
\begin{equation}
    \label{eq:hyb}
    E^\mathrm{hyb}_\mathrm{xc}=b E^\mathrm{HF}_\mathrm{x}+(1-b) E^\mathrm{GGA}_\mathrm{x}+E^\mathrm{GGA}_\mathrm{c},
\end{equation}
where the coefficient $b$ is a mixing parameter, and the labels "x" and "c" correspond to the exchange and correlation energy, respectively.
The energies $E^\mathrm{GGA}_\mathrm{x}$ and $E^\mathrm{GGA}_\mathrm{c}$ are calculated using the electron density and its gradient. 
A calculation of $E^\mathrm{HF}_\mathrm{x}$ involves occupied Kohn-Sham orbitals and thus depends on the electron density $\rho$ implicitly.
$E^\mathrm{hyb}_\mathrm{xc}[\rho]$, thus, is known as a hybrid functional.

Computational studies employ numerous hybrid functionals with a more general expression for $E^\mathrm{hyb}_\mathrm{xc}[\rho]$ than Eq.~\ref{eq:hyb}, since the LDA exchange and correlation energies (e.g., in the B3LYP functional) as well as the screened or range-separated exchange can contribute to the mix.
A constant part in all hybrid functionals is the orbital-dependent exchange term that makes this type of a calculation typically much more expensive than GGA.

In comparison to GGA, hybrids also present additional numerical challenges.
For instance, the integrable $\mathbf{q}=0$ singularity \cite{Gygi1986} leads to a slow convergence of total and band energies. 
Techniques for avoiding it are diverse with different performance  \cite{Massida1993,Spencer2008,Betzinger2010,Holzwarth2011}.
Another issue is the influence of basis sets or pseudopotentials that are typically designed for (semi)local functionals and nevertheless used with hybrids.
A recent study \cite{Borlido2020} discussed the numerical effect of this mismatch, where a pseudopotential was generated with one functional and applied in calculated with another one.
They found that such an inconsistency leads to an error in band gaps up to $\sim 1$~eV.
In other words, hybrid functionals introduce additional sources of errors compared to GGA.
A reproducibility study in Ref.~\cite{Lejaeghere2016} gave an optimistic view that modern electronic-structure codes with very different implementations yield essentially the same results (within experimental uncertainty) for elemental solids.
This study used the PBE functional, and it took a directed effort to arrive at this result.
Furthermore, there is a room for questions on reproducibility across the codes for systems beyond elemental solids.
It is natural then to expect that reaching reproducibility on the level of hybrid functionals presents a steeper challenge.
In an attempt to apply a superior method to GGA, one may resort to hybrids regardless of high computational expenses, and, then, it is important to be sure that such a calculation is not plagued by numerical errors.

What level of precision is needed in DFT calculations?
A realistic requirement for a numerical error in a comparison of atomization energies or similar quantities with experiment is chemical accuracy (1~kcal/mol$\approx$1.6~mHa).
For electronic transition and ionization energies, it is typically ~50--100~meV.
In high-quality benchmarks, the precision requirements become more stringent. 
We define the precision targets as 0.1~kcal/mol for atomization energies and 10--20~meV for transition energies from DFT calculations.

In the light numerical uncertainties due to the Fock exchange, it is important to have a highly precise tool that can be used for reference calculations or setting benchmarks.
The \textit{gold standard} for solids is the linearized-augmented plane waves (LAPW) method, and recent studies used it, for instance, as a reference tool for validating pseudopotentials \cite{Hamann2013,Garrity2014,vansetten2018,Borlido2020} and fully relativistic calculations in the framework of the numerical atomic orbitals \cite{Huhn2017}.
Indeed, it was shown that LAPW can attain absolute total energies within 1--2~$\mu$Ha/atom of the exact limit in calculations with (semi)local functionals \cite{Gulans2018}.
Although reaching the 1-$\mu$Ha limit is unlikely to be necessary in practical calculations, such a demonstration is important for showing how the LAPW basis can be improved systematically and that the DFT implementation works correctly.

Previous studies reported on implementations of hybrid functionals in LAPW~\cite{Betzinger2010,Tran2011}.
The approach in Ref.~\cite{Tran2011} employs the second variation. 
It means that, at every self-consistency step, the Hamiltonian is constructed and diagonalized for a local potential, and a few so-obtained KS orbitals are used as the basis for the non-local problem.
The method described in Ref.~\cite{Betzinger2010} maps the exchange matrix calculated for a few KS orbitals onto the entire LAPW basis, and does not involve a two-step diagonalization procedure. 
Despite the differences, results obtained with both methods depend on the number of used orbitals. 
This point makes handling precision tedious in high-throughput calculations.

In this study, we overcome the described issues with hybrids in LAPW by implementing an alternative approach. 
It employs the adaptively compressed exchange (ACE) operator~\cite{Lin2016} and is new to the LAPW framework.
We show that ACE yields results identical to the exact Fock exchange without numerical approximations. 
Our discussion is restricted to the PBE0 hybrid functional defined with the fraction $b=0.25$ of the $E^\mathrm{HF}_\mathrm{x}$ \cite{Perdew1996a}.
The implementation is open source and is available in the full-potential all-electron package \texttt{exciting} \cite{Gulans2014a}.

This paper is structured as follows.
In Sec.~\ref{sec:basis}, we give an overview to the LAPW basis complemented with local orbitals (LO) and explain how we apply it in calculations with the Fock exchange.
In Sec.~\ref{sec:matrix} and \ref{sec:ace}, we describe our implementation of the matrix elements and ACE in the LAPW formalism.
Sec.~\ref{sec:ace} contains also a discussion of similarities between the low-rank approximation in ACE and the implementation of the Fock exchange in Ref.~\cite{Betzinger2010}.
In Sec.~\ref{sec:atoms}, we calculate the total energies of a few atoms and molecules using GGA and hybrid exchange-correlation functionals.
The results are verified in a comparison to the data obtained with the multi-resolution analysis (MRA) in Ref.~\cite{Jensen2017}.
Sec.~\ref{sec:periodic} contains a benchmark calculation of transition energies in selected semiconductors and insulators.
All data obtained in this study are available in an open-access repository\cite{Gulans2022data}.

\section{LAPW+LO basis \label{sec:basis}}

The LAPW+LO basis consists of two components: LAPWs and local orbitals(LOs).
LAPWs are defined as
\begin{equation}
\label{eq:lapw}
    \phi_\mathbf{G+k}(\mathbf{r})=\left\{ 
    \begin{array}{lc}
       \sum\limits_{s \ell m} A^\mathbf{G+k}_{s \ell m} u_{s\ell}(r;E_{\ell}) Y_{\ell m}(\hat{r}) & r\in \mathcal{MT} \\
       \frac{1}{\sqrt{\Omega}}e^{i(\mathbf{G+k})\mathbf{r}} & r \in \mathcal{I}
    \end{array},
    \right.
\end{equation}
where the functions $u_{s\ell}(r;E_{\ell})$ represent $u_{\ell}(r;E_{\ell})$, solutions of the radial Sch\"{o}dinger equation, and their energy-derivatives $\dot{u}_{\ell}(r;E_{\ell})$~\cite{Slater1937}.
The index $s$ runs through the radial functions corresponding to the same spherical harmonic $Y_{lm}(\hat{r})$. 
The coefficients $A^\mathbf{G+k}_{s \ell m}$ ensure a smooth matching between the muffin-tin and interstitial regions denoted as $\mathcal{MT}$ and $\mathcal{I}$, respectively.
The radial Schr\"{o}dinger equation
\begin{equation}
\label{eq:radial}
    -\frac{1}{2r^2}\frac{\partial}{\partial r}\left(r^2 \frac{\partial u_{\ell}}{\partial r} \right) + \left( \frac{\ell(\ell+1)}{2r^2}+v_\mathrm{s}(r)- E_{\ell} \right) u_{\ell} = 0
\end{equation}
is solved with a predefined choice of the energy parameter $E_{\ell}$ and contains $v_\mathrm{s}(r)$, which is the spherically symmetric component of the local Kohn-Sham (KS) potential.

Local orbitals (LO) are defined as
\begin{equation}
\label{eq:lo}
    \phi_\nu(\mathbf{r})=\left\{ 
    \begin{array}{lc}
       \sum\limits_{s} a_{s} u_{s \ell}(r;E_{s\nu}) Y_{\ell m}(\hat{r}) & r\in \mathcal{MT} \\
       0 & r \in \mathcal{I}
    \end{array}
    \right. ,
\end{equation}
where the coefficients $a_{s}$ are chosen such that LOs are normalized and decay to zero\cite{Sjostedt2000}. 
 

To illustrate how LAPW+LO basis works in the $\mathcal{MT}$ region in a case of fully local KS potential, we consider a Be atom as an example. 
This system has two occupied orbitals: semicore $\psi_{1s}(\mathbf{r})$ and valence $\psi_{2s}(\mathbf{r})$ with the energies $\varepsilon_{1s}$ and $\varepsilon_{2s}$, respectively.
Since $v_\mathrm{s}(r)$ in Eq.~\ref{eq:radial} matches the KS potential of the full problem, the radial functions $u_0(r;E_{1s})$ and $\dot{u}_0(r;E_{1s})$ form an excellent basis for representing $\psi_{1s}(\mathbf{r})$ provided that $E_{1s}\approx\varepsilon_{1s}$.
Up to a prefactor, a Taylor expansion over the energy yields
\begin{equation}
\label{eq:linearization}
 \psi_{1s}(\mathbf{r})=u_0(r;E_{1s})+(\epsilon_{1s}-E_{1s})\dot{u}_0(r;E_{1s})+\dots
\end{equation}
and therefore shows that $u_0(r;E_{1s})$ and $\dot{u}_0(r;E_{1s})$ are the most important degrees of freedom for representing $\psi_{1s}(\mathbf{r})$.
Taking into account similar considerations for $\psi_{2s}(\mathbf{r})$, we conclude that also $u_0(r;E_{2s})$ and $\dot{u}_0(r;E_{2s})$ need to appear in the LAPW+LO basis as individual degrees of freedom.
To fulfill these requirements, we choose $E_0=E_{2s}$ for the LAPWs and the following LOs: 
\begin{equation}
\begin{array}{ll}
\label{eq:loset}
    \phi_1(\mathrm{r}) & = a_1 u_s(r;E_{1s})+b_1\dot{u}_s(r;E_{1s}),  \\
    \phi_2(\mathrm{r}) & = a_2 u_s(r;E_{1s})+b_2 u_s(r;E_{2s}),  \\
    \phi_3(\mathrm{r}) & = a_3 u_s(r;E_{2s})+b_3\dot{u}_s(r;E_{2s}),  \\
\end{array}
\end{equation}
where $a_i$ and $b_i$ are coefficients that satisfy the normalization and boundary conditions.
An additional LO
\begin{equation}
\label{eq:extralo}
    \phi_4(\mathrm{r}) = a_4 \dot{u}_s(r;E_{1s})+b_4\ddot{u}_s(r;E_{1s})
\end{equation}
makes this LO setup even more precise and less sensitive to a particular choice of $E_{1s}$.
It introduces an extra degree of freedom $\ddot{u}_s(r;E_{1s})$ that corresponds to a quadratic term in Eq.~\ref{eq:linearization}.
Such a strategy yields an LAPW+LO basis sufficient for microhartree precision\cite{Gulans2018} in calculations with local functionals.
We label the LO sets constructed according to these lines as \texttt{stdlo}.

In the case of a hybrid functional, it is not clear how to obtain radial functions consistent with a non-local KS potential. 
We follow the usual approach where the radial functions are generated using a local potential in Eq.~\ref{eq:radial} although it is inconsistent with the hybrid calculation.
Such an approach makes the expansion in Eq.~\ref{eq:linearization} invalid, since the left- and the right-hand sides correspond to different Hamiltonians.
Therefore, it cannot be taken for granted that the LO set defined in Eqs.~\ref{eq:loset} and \ref{eq:extralo} ensure convergence of total energies and band energies. 
In this study, we solve this problem by adding high-energy LOs as much as necessary. 
They are constructed as a linear combination of $u_\ell(r;E_{n\ell})$ and $\dot{u}_\ell(r;E_{n\ell})$, where the energy parameters $E_{n\ell}$ are chosen such that the next LO has an extra node in the $\mathcal{MT}$ region compared to the previous one.
Following Ref.~\cite{Michalicek2013}, we refer to LOs constructed in this manner as high-energy local orbitals (\texttt{helo}).

\section{Matrix elements \label{sec:matrix}}
We consider the Fock exchange operator defined as
\begin{equation}
\label{eq:vxl}
 \hat{V}_\mathrm{x}\psi_{n\mathbf{k}}(\mathbf{r})=-\sum\limits_{a\mathbf{k}^\prime}\psi_{a\mathbf{k}^\prime}(\mathbf r)\int\frac{\psi_{a\mathbf{k}^\prime}^\ast(\mathbf{r}^\prime) \psi_{n\mathbf{k}}(\mathbf{r}^\prime)}{|\mathbf{r}-\mathbf{r}^\prime|}\,\dd \mathbf{r}^\prime ,
\end{equation}
and its matrix elements
\begin{equation}
\label{eq:matrixelements}
M^{\mathbf{k}}_{nn^\prime}=\langle \psi_{n\mathbf{k}} |\hat{V}_\mathrm{x}|\psi_{n^\prime\mathbf{k}}\rangle .
\end{equation}
The summation indices $a$ and $\mathbf{k}^\prime$ run over all occupied bands including dispersionless core orbitals and reducible $\mathbf{k}$-points in the Brillouin zone, respectively.
In the present work, we employ either the non-relativistic or scalar-relativistic approximation for the valence wavefunctions.
However, the core orbitals are obtained from the four-component radial Dirac equation. 
This mismatch breaks degeneracy of bands in symmetric systems whenever the core contains $p$-orbitals.
To circumvent this issue, we follow the approach described in Ref.~\cite{BetzingerThesis}. 
For each core shell, we use weighted average over the available $j$ values according to their multiplicity $2j+1$.

We evaluate the matrix elements in four steps following Eqs.~\ref{eq:vxl} and \ref{eq:matrixelements} directly. 
(i) Calculate overlap charge densities literally as
\begin{equation}
\label{eq:density}
\rho_{{a\mathbf{k}^\prime},{n\mathbf{k}}}(\mathbf{r})=\psi_{a\mathbf{k}^\prime}^\ast(\mathbf{r}) \psi_{n\mathbf{k}}(\mathbf{r}).
\end{equation}
We employ fast Fourier transform (FFT) grids for calculating the $\mathcal{I}$ term and the Lebedev grids for $\mathcal{MT}$ term.
(ii) Evaluate the electrostatic potential $v_{{a\mathbf{k}^\prime},{n\mathbf{k}}}(\mathbf{r})$ due to the density $\rho_{{a\mathbf{k}^\prime},{n\mathbf{k}}}(\mathbf{r})$ using the pseudocharge method\cite{Weinert1981}.
(iii) Calculate 
\begin{equation}
\label{eq:w}
w_{n\mathbf{k}}(\mathbf{r})=-\sum\limits_{a\mathbf{k}^\prime} \psi_{a\mathbf{k}^\prime}(\mathbf{r}) v_{{a\mathbf{k}^\prime},{n\mathbf{k}}}(\mathbf{r}),
\end{equation}
and, similarly to step (i), we perform this operation using the FFT and Lebedev grids.
(iv) Calculate the matrix elements as
\begin{equation}
    M^{\mathbf{k}}_{nn^\prime}=\int 
    \psi^\ast_{n\mathbf{k}}(\mathbf{r}) w_{n^\prime\mathbf{k}}(\mathbf{r})
    \dd\mathbf{r}.
\end{equation}


The described procedure is similar to the one described in Ref.~\cite{Massida1993}, but there are a few important differences.
Ref.~\cite{Massida1993} skips step (iii) and evaluates the matrix elements as
\begin{equation}
    M^{\mathbf{k}}_{nn^\prime}=\sum\limits_{a\mathbf{k}^\prime}\int 
    \rho_{{n\mathbf{k}},{a\mathbf{k}^\prime}}(\mathbf{r})v_{{a\mathbf{k}^\prime},{n^\prime\mathbf{k}}}(\mathbf{r})
    \dd\mathbf{r}.
\end{equation}
Such an approach is not suitable for our purposes, because the ACE formalism requires that the quantity $\hat{V}_\mathrm{x}\psi_{n\mathbf{k}}(\mathbf{r})$ is available.
Furthermore, the procedure described here requires $O( N^2_\mathrm{v} N_\mathrm{LAPW} \log N_\mathrm{LAPW})$ floating point operations, whereas the algorithm from Ref.~\cite{Massida1993} formally scales as $O( N^3_\mathrm{v} N_\mathrm{LAPW})$, where $N_\mathrm{v}$ is the number of the valence bands.

An alternative approach defines an auxiliary basis (also known as a mixed basis) for representing products of wave functions \cite{Betzinger2010}.   
Its implementations in Refs.~\cite{Betzinger2010,Gulans2014a} scale roughly as $O(N_\mathrm{v}^2 N_\mathrm{LAPW}^2)$ in terms of floating-point operations and require a storage of $O(N_\mathrm{LAPW}^2)$ Coulomb matrix elements. 
This approach is well-suited for close-packed systems, but it requires a revision for sparse and molecular systems.
As our scope goes beyond close-packed systems, we do not employ the product-basis formalism in the present work.  

The integral in Eq.~\ref{eq:vxl} diverges when $\mathbf{q}=\mathbf{k}^\prime-\mathbf{k}$ is equal to 0 and $a=n$.
This singularity is integrable in the $\mathbf{q}$-space, and there are several practical approaches to handle it properly~\cite{Massida1993,Spencer2008,Betzinger2010,Sundararaman2013}.
In this study, we use (i) the auxiliary function method as described in Ref.~\cite{Massida1993} and (ii) a Poisson equation solver with the free boundary conditions \cite{Goedecker1998,Fisicaro2016}. 
In the option (i), one selects a function $F(\mathbf{q})$ such that it diverges at $\mathbf{q}=0$ with the same asymptotic behavior as the integral in Eq.~\ref{eq:vxl}.
It enables a correction that cancels the singularity exactly in the diagonal matrix elements for valence bands:
\begin{equation}
    \label{eq:mpb}
    \begin{split}
    M^{\mathbf{k}}_{aa}=&
    -\sum\limits_{\mathbf{q}} 
   \sum\limits_{a^\prime} \iint\frac{\rho_{{a\mathbf{k}},{a^\prime\mathbf{k}+\mathbf{q}}}(\mathbf{r})\rho_{{a^\prime\mathbf{k}+\mathbf{q}},{a\mathbf{k}}}(\mathbf{r})}{|\mathbf{r}-\mathbf{r}^\prime|}
    \dd\mathbf{r}\dd\mathbf{r^\prime}  \\
   & + \frac{1}{N_\mathbf{k}} \sum\limits_{\mathbf{q}} 
    F(\mathbf{q}) - \frac{\Omega}{(2\pi)^3}\int\limits_\mathrm{BZ} F(\mathbf{q}) \dd \mathbf{q}.
   \end{split}
\end{equation}
We follow the choice of the auxiliary function proposed by Massida \textit{et al.}~\cite{Massida1993} and set
\begin{equation}
    \label{eq:auxf}
    F(\mathbf{q})=\frac{4\pi}{\Omega}\sum\limits_{\mathbf{G}+\mathbf{q}} \frac{e^{-\beta(\mathbf{G}+\mathbf{q})^2 }}{|\mathbf{G}+\mathbf{q}|^2},
\end{equation}
where $\beta$ is a parameter that we choose explicitly depending on a considered system.
Eqs.~\ref{eq:mpb} and \ref{eq:auxf} essentially define a correction that we denote as $\Delta$.
To apply it in practice, the first two terms in Eq.~\ref{eq:mpb} are evaluated ignoring the $\mathbf{G+q}=0$ component.
The remaining term is evaluated analytically.

The auxiliary function method reduces to evaluating
\begin{equation}
\label{eq:q0corr}
M^{\mathbf{k}}_{nn^\prime}=\langle \psi_{n\mathbf{k}} |\hat{V}_\mathrm{x}|\psi_{n^\prime\mathbf{k}}\rangle_\mathrm{nd} + \Delta_{n\mathbf{k}}\sum\limits_{a}\delta_{an}
\end{equation}
with precomputed $\Delta_{n\mathbf{k}}$ and the subscript "nd" that shows that the $\mathbf{G+q}=0$ component is ignored.
This correction has to be applied not only to the matrix $M^{\mathbf{k}}$, but also to the functions $\hat{V}_\mathrm{x}\psi_{n\mathbf{k}}(\mathbf{r})$ required in the ACE method. 
We make $M^{\mathbf{k}}$ and $\hat{V}_\mathrm{x}\psi_{n\mathbf{k}}(\mathbf{r})$ consistent, by applying the following correction
\begin{equation}
\label{eq:q0corrw}
\hat{V}_\mathrm{x}|\psi_{n\mathbf{k}}\rangle=\hat{V}_\mathrm{x}|\psi_{n\mathbf{k}}\rangle_\mathrm{nd} + \Delta_{n\mathbf{k}}\sum\limits_{a}\delta_{an}|\psi_{a\mathbf{k}}\rangle.
\end{equation}

As discussed in Ref.~\cite{Paier2005}, ignoring the divergent terms without applying a $\mathbf{q}=0$ correction leads to an error that decays as $O(1/\Omega^{1/3})$ for isolated systems. 
When the correction is applied, the error scales as $O(1/\Omega)$. 
In bulk systems, we vary the BZ zone sampling rather than the volume of the unit cell. 
A calculation of a primitive unit cell with $N_\mathbf{k}=N_1 \times N_2 \times N_3$ $\mathbf{k}$-points corresponds to a $\Gamma$-point calculation of a supercell consisting of $N_1 \times N_2 \times N_3$ primitive unit cells.
Therefore, we anticipate that the error of the order $O(1/\Omega)$ in $\Gamma$-only calculations of isolated systems translates into the error $O(1/N_\mathbf{k})$ for periodic systems. 
We use this reasoning and extrapolate band energies and total energies to the converged limit. This approach allows us to reach the targeted level of precision, although Refs. \cite{Spencer2008,Sundararaman2013,Duchemin2010} have shown that it is possible to achieve even faster convergence rate with respect to $N_\mathbf{k}$.

In the option (ii), we apply a multi-wavelet Poisson solver~\cite{Fisicaro2016}. 
In a nutshell, it expresses the Coulomb kernel $1/|\mathbf{r}-\mathbf{r}^\prime|$ as a linear combination of Gaussian functions.
Such a representation allows one to separate the three-dimensional Coulomb integrals into three one-dimensional convolutions, which are evaluated using wavelets.
This method supports the boundary conditions suitable for calculations with periodicity in zero to three directions. 
We use the implementation in the PSolver library supplied with the BigDFT package\cite{Genovese2006}.
To make it work with LAPW, we follow the Weinert's method \cite{Weinert1981} for calculating the electrostatic potential and apply the wavelet-based solver for the pseudocharge density.

\section{Fock exchange in low-rank representation \label{sec:ace}}


We approximate the exchange operator with a low-rank approximation
\begin{equation}
  \label{eq:aceproj}
    \hat{V}^\mathrm{ACE}_\mathrm{x} = -\sum\limits_{n\mathbf{k}} |\xi_{n\mathbf{k}}\rangle\langle\xi_{n\mathbf{k}}|,
\end{equation}
where $\xi_{n\mathbf{k}}(\mathbf{r})$ are projector functions chosen such that $\hat{V}^\mathrm{ACE}_\mathrm{x}|\psi_{n\mathbf{k}}\rangle=\hat{V}_\mathrm{x}|\psi_{n\mathbf{k}}\rangle$ for a range of orbitals covering all occupied and a few lowest unoccupied states. 
As chosen in Ref.~\cite{Lin2016}, this property is satisfied if the exchange operator is approximated by
\begin{equation}
\label{eq:ace2}
    \hat{V}^\mathrm{ACE}_\mathrm{x} = \sum\limits_{nn^{\prime}\mathbf{k}} | w_{n\mathbf{k}} \rangle 
    \left( M^{-1} \right)_{nn^\prime} \langle w_{n^\prime\mathbf{k}}|.
\end{equation}
For the sake of compactness, the index $\mathbf{k}$ is omitted from $M^\mathbf{k}$ and related matrices introduced below.
Eq.~\ref{eq:ace2} reduces to Eq.~\ref{eq:aceproj} after the factorization $M^{-1}=-B^\dagger B$.
Since $M$ is negative defined, it can be expressed via the Cholesky decomposition as $ M=-L L^\dagger$, where $L$ is a lower triangular matrix.
Recognizing that $B=L^{-1}$, one arrives at
\begin{equation}
  \xi_{n\mathbf{k}}(\mathbf r)=\sum\limits_{n^\prime}  B_{nn^\prime} w_{n^\prime\mathbf{k}}(\mathbf{r}).
\end{equation}

In our implementation, we store $\xi_{n\mathbf{k}}(\mathbf r)$ similarly to other quantities in the LAPW formalism, i.e., considering the $\mathcal{MT}$ and $\mathcal{I}$ contributions separately:
\begin{equation}
\label{eq:xirep}
  \xi_{n\mathbf k}(\mathbf r)=\left\{\begin{aligned}
  &\sum\limits_{lm} \xi^{n\mathbf k}_{lm}(r) Y_{lm}(\hat r)&&\mathbf r\in \mathcal{MT}\\
  &\xi_{n\mathbf G + \mathbf k}e^{i(\mathbf G + \mathbf k)\mathbf r}&&\mathbf r\in \mathcal{I}\end{aligned}\right. .
\end{equation}
A Fourier transform of the coefficients $\xi_{n\mathbf G + \mathbf k}$ yields a smooth function which we denote as $\bar{\xi}_{n\mathbf k}(\mathbf r)$.
This function coincides with $\xi_{n\mathbf k}(\mathbf r)$ in the $\mathcal{I}$ region.

To apply ACE in practical calculations, we derive expressions for the matrix elements with respect to the basis functions. 
The main ingredient is the overlap integral $P^\mathbf{k}_{n\beta}=\bra{\xi_{n\mathbf k}}\ket{\phi_\beta}$, where $\beta$ is a unified basis-function index that covers all LAPWs and LOs at a given $\mathbf{k}$-point.
In the case of LAPWs, $\bra{\xi_{n\mathbf k}}\ket{\phi_{\mathbf G+\mathbf k}}$ splits into contributions from the $\mathcal{MT}$ and $\mathcal{I}$ regions:
\begin{equation}
\label{eq:ximat}
    \bra{\xi_{n\mathbf k}}\ket{\phi_{\mathbf G+\mathbf k}}=
    \bra{\xi_{n\mathbf k}}\ket{\phi_{\mathbf G+\mathbf k}}_{\mathcal{I}}+
    \bra{\xi_{n\mathbf k}}\ket{\phi_{\mathbf G+\mathbf k}}_{\mathcal{MT}}.
\end{equation}
The first term is expressed as
\begin{equation}
\label{eq:xilapw}
    \braket{\xi_{n\mathbf k}}{\phi_{\mathbf G+\mathbf k}}_{\mathcal{I}}=
    \frac{1}{\sqrt{\Omega}}\int\limits_\Omega \bar{\xi}^\ast_{n\mathbf k}(\mathbf r)\theta(\mathbf r)e^{i(\mathbf G+\mathbf k)\cdot\mathbf r}\, \dd \mathbf r ,
\end{equation}
where $\theta(\mathbf r)$ is a step function.
In practice, $\braket{\xi_{n\mathbf k}}{\phi_{\mathbf G+\mathbf k}}_{\mathcal{I}}$ is evaluated by performing a Fourier transform of the function $\bar{\xi}^\ast_{n\mathbf k}(\mathbf r)\theta(\mathbf r)$. 
In the second term of Eq.~\ref{eq:ximat}, we obtain
\begin{equation}
   \braket{\xi^{n\mathbf k}}{\phi_{\mathbf G+\mathbf k}}_{\mathcal{MT}}=\sum\limits_{s\ell m} A^{\mathbf G+\mathbf k}_{s\ell m}\int\limits_{0}^{R_\mathrm{MT}}
   \xi_{\ell m}^{n\mathbf k \ast}(r) u_{s \ell}(r) r^2\,\dd r.
\end{equation}

For the matrix elements with LOs, we have only the $\mathcal{MT}$ part which can be calculated as
\begin{equation}
\label{eq:xilo}
   \braket{\xi_{n\mathbf k}}{\phi_{\nu}}=\delta_{ll_\nu}\delta_{mm_\nu}\int\limits_{0}^{R_\mathrm{MT}}
   \xi_{lm}^{n\mathbf k*}(r) f_{\nu}(r) r^2\,\dd r .
\end{equation}
$\delta_{ll_\nu}$ and $\delta_{mm_\nu}$ enter the equation above to acknowledge that only one spherical harmonic enters a given LO. 

After the overlaps with the projectors are calculated, we express the contribution of the Fock exchange to the Hamiltonian matrix as
\begin{equation}
\label{eq:acematrix}
    \bra{\phi_{\alpha\mathbf{k}}} \hat{V}^\mathrm{ACE}_\mathrm{x}\ket{\phi_{\beta\mathbf{k}}}=
    -\sum\limits_{n}P^{\mathbf{k}\ast}_{n\alpha}P^{\mathbf{k}}_{n\beta},
\end{equation}

If the Hamiltonian is diagonalized directly, the matrix in Eq.~\ref{eq:acematrix} is constructed via matrix-matrix multiplication $V^\mathrm{ACE}_\mathrm{x}=P^\dagger P$.
However, the approach described in this section is suitable for iterative eigensolvers too.
In this case, we apply $\hat{V}^\mathrm{ACE}_\mathrm{x}$ on a trial wave function $\varphi(\mathbf{r})=\sum_\alpha C_\alpha \phi_\alpha(\mathbf{r})$ without constructing the matrix in Eq.~\ref{eq:acematrix} explicitly. 
We achieve this by choosing the following order of operations: $V^\mathrm{ACE}_\mathrm{x}C = P^\dagger (PC)$.

We note that ACE is not the only known possibility for expressing the exchange operator in the low-rank form of Eq.~\ref{eq:aceproj}.
In earlier work\cite{Betzinger2010}, Betzinger \textit{et al.} suggested the following approximation (which we label as BFB):
\begin{equation}
\label{eq:bfb}
    \hat{V}^\mathrm{BFB}_\mathrm{x} = \sum\limits_{nn^\prime\mathbf{k}}  
    | \psi_{n\mathbf{k}} \rangle\langle \psi_{n\mathbf{k}} |
    \hat{V}_\mathrm{x}  | \psi_{n^\prime\mathbf{k}} \rangle\langle \psi_{n^\prime\mathbf{k}} | .
\end{equation}
This approach is implemented in two LAPW codes~\cite{fleur,Gulans2014a} where it is used for hybrid calculations.
With the notation introduced in this section, Eq.~\ref{eq:bfb} reduces to Eq.~\ref{eq:aceproj} if
\begin{equation}
\label{eq:bfbproj}
  \xi_{n\mathbf{k}}(\mathbf r)=\sum\limits_{n^\prime}  L_{n^{\prime}n}^{\ast} \psi^\ast_{n^\prime\mathbf{k}}(\mathbf{r}).
\end{equation}
To apply this approach in the present study, we calculate the projector matrix $P$ and use it as in Eq.~\ref{eq:acematrix}.
The previous implementations, however, implement the Fock exchange contribution to the Hamiltonian as 
\begin{equation}
\label{eq:bfbmatrix}
    V^\mathrm{BFB}_\mathrm{x}\approx SZMZ^\dagger S,
\end{equation}
where $S$ is the overlap matrix of the basis functions and $Z$ is the matrix of wave function coefficients defined by $\psi_{n\mathbf{k}}(\mathbf{r})=\sum_\alpha Z_{\alpha n\mathbf{k}}\phi_{\alpha}(\mathbf{r})$.
Eq.~\ref{eq:bfbmatrix} reduces to Eq.~\ref{eq:acematrix} with $P=L^{\dagger}Z^\dagger S$.

The two low-rank approximations in Eqs.~\ref{eq:aceproj} and \ref{eq:bfb} serve the same purpose.
They map the exchange matrix obtained for a small set of orbitals onto a matrix for the entire basis set.
The two approaches are not equivalent due to a different choice of the projector functions.
However, BFB and ACE yield the same answer if one chooses to calculate matrix elements for all possible orbitals with a given basis.

Both considered low-rank approximations are used in a two-level self-consistency procedure\cite{Betzinger2010,Lin2016}. 
The inner cycle updates the electron density and the local part of the potential, whereas the outer cycle updates the non-local exchange operator.  
Such an approach allows one to reduce the number of  evaluations of the matrix $M^{\mathbf{k}}$ while increasing the total number of diagonalizations and local potential updates.

\section{\label{sec:atoms}Atoms and molecules}

Calculations in Ref.~\cite{Lin2016} showed that ACE yields the correct exchange energy even if $\hat{V}^\mathrm{ACE}_\mathrm{x}$ is constructed using only valence orbitals.
This result was obtained in a plane-wave pseudopotential calculation, and here we verify whether it remains so in an all-electron calculation.
For a comparison, we also check how the number of unoccupied orbitals taken along with the occupied ones influences the  total energy within the BFB approximation.

\begin{figure}
\includegraphics[width=8.5cm]{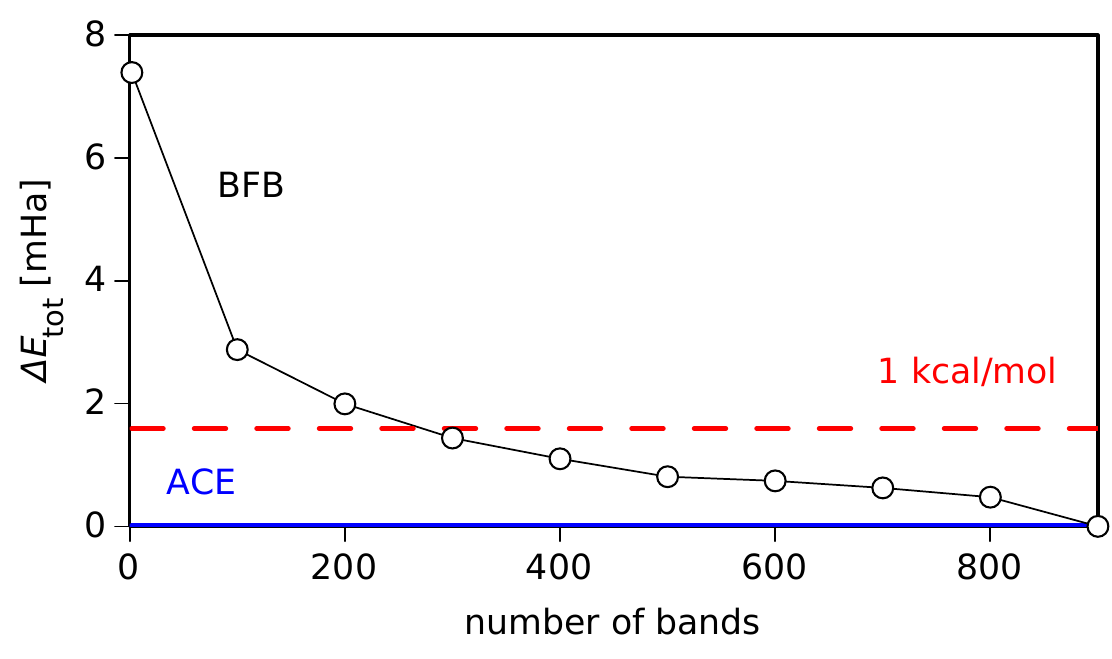}
\caption{\label{fig:bfb} 
Total energy of the Be atom with respect to the number of bands. 
The acronyms ACE and BFB label the exchange operators approximated via Eqs.~\ref{eq:aceproj} and \ref{eq:bfb}, respectively.
The red dashed line shows the limit of chemical accuracy.}
\end{figure}

We explore these two methods in a PBE0 calculation of the Be atom in a cubic unit cell with the dimension of $7$~bohr.
Figure~\ref{fig:bfb} shows the dependence on the number of bands $N_\mathrm{b}$ used for calculating the exchange matrix $M^\mathbf{k}$.
We observe that in the case of the ACE representation the total energy converges to the same value within 0.1~$\mu$Ha regardless whether $N_\mathrm{b}$ is set to 2 or 900.
The latter number corresponds to the maximum possible number with the employed basis settings, and the approximate exchange representation becomes exact.
The BFB calculation, in turn, shows that the self-consistent total energy slowly approaches the value obtained with ACE as $N_\mathrm{b}$ is increased.
It is required to set $N_\mathrm{b}\approx 300$ (one third of the all possible bands) to reach the precision level that corresponds to the chemical accuracy limit.

Typically the quantities of interest are energy differences rather absolute values.
It allows one to hope for a cancellation of errors.
However, Be is a light element with just 4 electrons, and the BFB errors will increase in more difficult cases if the convergence with respect to $N_\mathrm{b}$ is not ensured. 
Since there is no such issue with ACE and it requires essentially equal computational effort compared to BFB in our implementation, we do not apply BFB in our further calculations.

A similar convergence test with respect to $N_\mathrm{b}$ was performed in Ref.~\cite{Tran2011}, where hybrid functionals with the Fock and screened exchange were implemented using the second-variational technique.
The calculated error showed similar trends and the order of magnitude as in the present work for BFB.

\begin{figure}
\includegraphics[width=8.0cm]{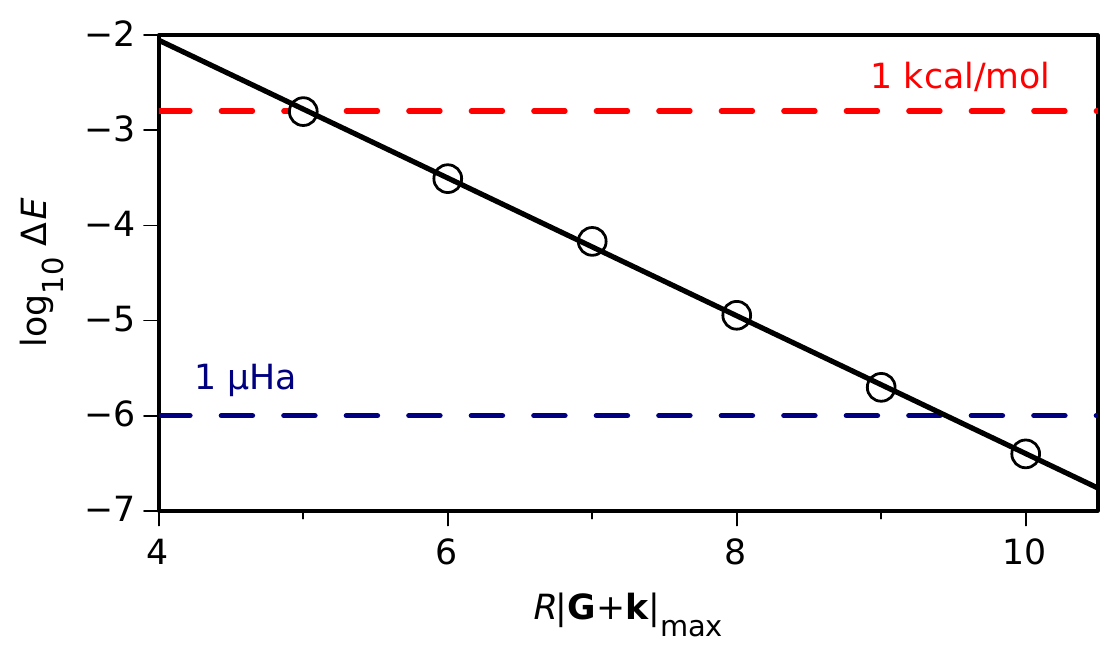}
\includegraphics[width=8.5cm]{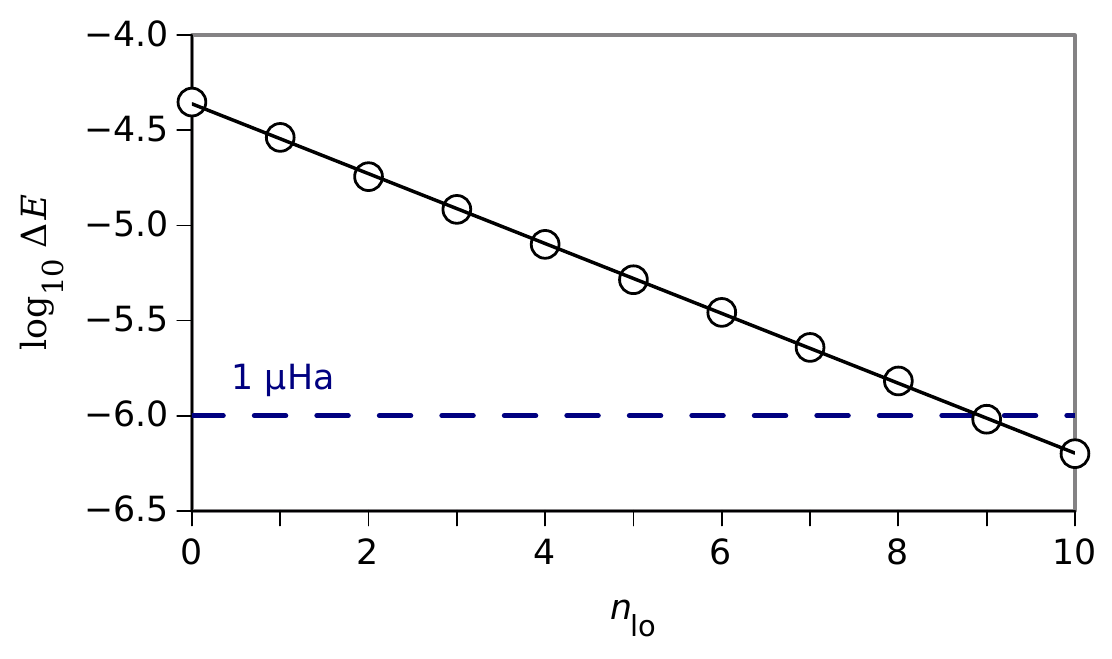}
\caption{\label{fig:Beconv} Error in the PBE0 total energy of a Be atom with respect to the LAPW cutoff (top) and the number of additional local orbitals (bottom).
The reference for the error calculation is the extrapolated value.}
\end{figure}

As the next step, we investigate the influence of the LAPW+LO parameters on the quality of the calculation.
The upper panel of Fig.~\ref{fig:Beconv} shows that the total energy converges exponentially with respect to the LAPW cutoff $R_\mathrm{MT}|\mathbf{G}+\mathbf{k}|_\mathrm{max}$.
The observed behavior is similar to what was observed in calculations with local functionals~\cite{Gulans2014a,Gulans2018} both, qualitatively and quantitatively. 
In particular, it takes $R_\mathrm{MT}|\mathbf{G}+\mathbf{k}|_\mathrm{max}\approx 10$ to reach the precision level of 1~$\mu$Ha, and the same threshold was obtained in Ref.~\cite{Gulans2018}.

As explained in Sec.~\ref{sec:basis}, we use the local PBE potential for constructing the radial functions in LAPWs and LOs, and it has an impact on the total energies.
The \texttt{stdlo} set consisting of four functions shown in Eqs.~\ref{eq:loset} and \ref{eq:extralo} is sufficient for reaching the complete-basis limit within 1~$\mu$Ha in a PBE calculation.
The same LO set results in an error of 44~$\mu$Ha when the PBE0 functional is employed. 
The lower panel of Fig.~\ref{fig:Beconv} shows the decay of the error in the total energy as \texttt{helo}'s (Eq.~\ref{eq:extralo}) are added.
We find that 9 of them are required to reach the precision threshold of 1~$\mu$Ha.
The error decays exponentially with the number of LOs in this calculation, but it is uncertain whether such a predictable behaviour can be generalized to other systems. 



Next, we compare the total energies of isolated non-relativistic atoms and molecules to the results obtained with the multi-resolution analysis in Ref.~\cite{Jensen2017,Jensen2017data}.
Geometries of the considered molecules are taken from the same reference.
Our calculations are performed using cubic unit cells sufficiently large to avoid an interaction of periodic images. 
The dimensions are $20$~bohr in all cases, except for, the Be atom for which $25$~bohr are required. 
The employed LAPW cutoffs are $R_\mathrm{MT}|\mathbf{G}+\mathbf{k}|_\mathrm{max}=$7--10 depending on the chemical element.
Such a choice leads to a large number of basis functions which exceeds $10^5$ for H$_2$, CH$_4$ and CO.
We solve these large eigenproblems using a modified Davidson eigensolver\cite{Gulans2018}.
Calculations for each system are performed with the PBE and PBE0 exchange-correlation functionals to verify that the LAPW cutoff is sufficiently large and that the interaction between the periodic neighbors is sufficiently weak.
The PBE calculations are performed with the \texttt{stdlo} settings. 
In the case of PBE0, we employ both, the \texttt{stdlo} and expanded sets of LOs. 
The latter ones are constructed taking \texttt{stdlo} as a starting point and expanding it with 8--12 \texttt{helo}'s. 
This approach is labeled below simply as \texttt{helo}.

\newcommand{\mc}[3]{\multicolumn{#1}{#2}{#3}}
\newcommand{\cw}{\columnwidth}
\begin{table}
\caption{\label{tab:atomsx} Comparison of PBE and PBE0 total energies between LAPW and MRA. LAPW energies are labeled with \texttt{stdlo} and \texttt{helo} and given relative to MRA energies.}
\newcolumntype{L}[1]{>{\raggedright\let\newline\\\arraybackslash\hspace{0pt}}m{#1}}
\newcolumntype{R}[1]{>{\raggedleft\let\newline\\\arraybackslash\hspace{-15pt}}m{#1}}
          
\begin{threeparttable}
    \setlength{\tabcolsep}{0.05cm} 
    \begin{tabularx}{\columnwidth}{@{\extracolsep{\fill}}lrrrrrrr}
        \tdrule
        & \mc{3}{c}{PBE} & & \mc{3}{c}{PBE0} \\[1pt]
        \cline{2-4}\cline{6-8}\\[-8pt]
        
        & 
        \mc{2}{l}{\texttt{stdlo}} & 
        \mc{1}{c}{MRA\tnote{1}} &
        &
        \mc{1}{c}{\texttt{stdlo}} & 
        \mc{1}{c}{\texttt{helo}} & 
        \mc{1}{c}{MRA\tnote{1}} \\
        
        & 
        \mc{2}{l}{$[\mathrm{\mu Ha}]$} & 
        \mc{1}{c}{$\mathrm{[Ha]}$} & 
        &
        \mc{1}{c}{$[\mathrm{\mu Ha}]$} & 
        \mc{1}{c}{$[\mathrm{\mu Ha}]$} & 
        \mc{1}{c}{$\mathrm{[Ha]}$} \\[3pt]
        
        \hline
        He  &  0.7 & &  -2.8929349  & & 42.4  &  0.9  &  -2.8951780  \\  
        Be  &  0.4 & &  -14.6299479  & &  45.3  &  2.2  &  -14.6366416  \\  
        H$_2$  &  2 & &  -1.166700  & &  4  &  3  &  -1.169064  \\
        CH$_4$  &  4 & &  -40.468109  & &  53  &  6  &  -40.478725  \\  
        CO  &  3 & &  -113.242609  & &  114  &  7  &  -113.238001  \\
        \tdrule
    \end{tabularx}
    \begin{tablenotes}
        \item [1] Reference \cite{Jensen2017}
    \end{tablenotes}
\end{threeparttable}

\end{table}

Tab.~\ref{tab:atomsx} shows a comparison of PBE and PBE0 total energies between LAPW and MRA.
All PBE energies obtained with the two methods agree up to 4~$\mu$Ha.
We find such a level of agreement remarkable and do not seek to improve it further. 
The obtained result shows that the LAPW+LO basis is sufficiently flexible, and the unit cells are large enough to avoid the interaction between the periodic images.

A comparison of PBE0 calculations shows that the  \texttt{stdlo} settings do not allow us to achieve the same level of precision in general.
Only in the case of the H$_2$ molecule, the difference between the LAPW and MRA energies is 3~$\mu$Ha.
For all other systems, it ranges between 42 and 115~$\mu$Ha. 
The reason why the H$_2$ stands out is the small $R_\mathrm{MT}=0.7$~bohr employed for the H atoms due to the short bond length in this molecule. 
In this view, the \texttt{stdlo} results should be considered with some caution, because they depend on the $\mathcal{MT}$ radii.
Once the \texttt{helo}'s are introduced, the agreement between LAPW and MRA improves and becomes similar to what is achieved in the PBE calculations. 

The results presented in Tab.~\ref{tab:atomsx} show that ACE provides a precise representation of the Fock exchange for total energy calculations. 
The main restriction in precision stems from the radial basis rather than this method as such.

\section{Solids \label{sec:periodic}}

We apply the ACE method in a calculation of the band structure for a set of semiconductors and insulators.
This set was compiled in Ref.~\cite{Paier2006} where hybrid functionals were implemented using the projector-augmented wave method. 
The same set was used later for benchmarking two LAPW implementations~\cite{Betzinger2010,Tran2011}. 
The published data consist of transition energies calculated with the PBE and PBE0 exchange-correlation functionals in three different codes and, thus, offer us excellent framework for an analysis. 
In contrast to the previous all-electron studies, our calculation does not involve the number of bands as a convergence parameter. 

\begin{figure}
\includegraphics[width=8.5cm]{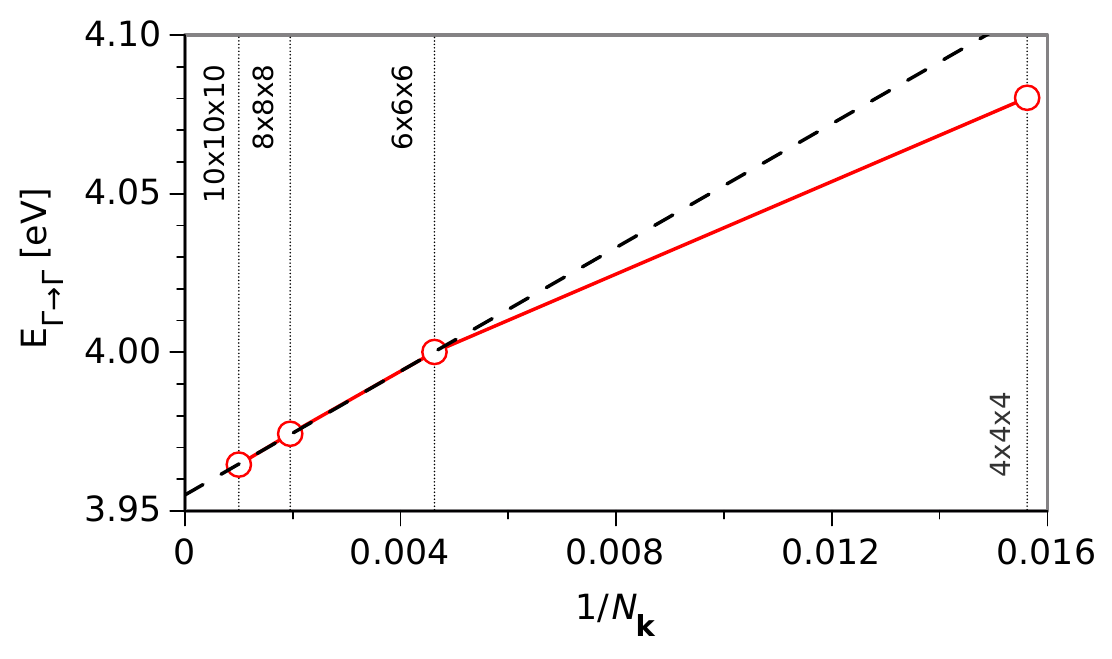}
\caption{\label{fig:Si_extrap} 
$\Gamma\rightarrow\Gamma$ transition energy of bulk silicon with respect to different $\mathbf{k}$-grids.
Dashed black line corresponds to a linear fit for three data points corresponding to the largest considered grids.}
\end{figure}

\sisetup{range-units=single, range-phrase=-}
As in previous studies, we use the experimental lattice geometries (see Ref.~\cite{Tran2011} for the specific parameters). 
The $\mathcal{MT}$ radii are chosen close to maximum possible values restricted by the interatomic distances. 
The chosen LAPW cutoff  $R_\mathrm{MT}|\mathbf{G}+\mathbf{k}|_\mathrm{max}$ is chosen such that it guarantees at least $\SIrange{1}{2}{meV}$ precision for $\Gamma \rightarrow \Gamma$ transition. 
In particular, we set $R_\mathrm{MT}|\mathbf{G}+\mathbf{k}|_\mathrm{max}=8$ for C and Si, 9 for Ar, 10 for MgO, and 11 for NaCl and GaAs. 
All calculations employ the scalar-relativistic zero-order relativistic approximation for the valence and semicore electrons\cite{vanLenthe1993}.
According to our estimates, the chosen LAPW cutoffs and the relativistic model yield a combined uncertainty of 4~meV.
Following the same reasoning as in Sec.\ref{sec:atoms}, we use two sets of LOs labeled as \texttt{stdlo} and \texttt{helo} for both, PBE and PBE0 calculations. 
Finally, core orbitals (Ar $1s^2$, Si $1s^2$, Ga $1s^22s^22p^6$, As $1s^22s^22p^6$, Mg $1s^2$, Na $1s^2$ and Cl $1s^22s^22p^6$) are obtained in a PBE calculation and remain fixed during the self-consistent PBE0 run. 
This approach is consistent with Refs.~\cite{Betzinger2010} and \cite{Tran2011}.

As explained in Sec.~\ref{sec:matrix}, we anticipate that the error after applying the $\mathbf{q}=0$ correction of the divergent term scales as $O(1/N_{\mathbf{k}})$. 
We apply this reasoning for extrapolating the PBE0 band energies to the converged limit.
Fig.~\ref{fig:Si_extrap} shows the calculated $\Gamma \rightarrow \Gamma$ transition energies in bulk Si with respect to $\mathbf{k}$-grid size $N_{\mathbf{k}}=N\times N\times N$, where $N=4,6,8,10$.
The results obtained with $N>4$ follow the anticipated asymptotic relation extremely well, and only the data point with $N=4$ does not follow this trend.
We find this behavior in all considered materials, and, therefore, perform the linear fit for transition energies calculated with $N=$6--10 and extrapolate these results to the limit of infinite $N_\mathbf{k}$. 
The quality of the linear fit convinces us that the error due to such a procedure is well under 1~meV.
In the PBE calculations, no extrapolation is needed and the energy gaps are sufficiently converged at $N_{\mathbf{k}}=10\times 10\times 10$ used in our calculations. 

\begin{table*}
\caption{Transition energies (in eV) of various solids calculated using PBE and PBE0 functionals.}\label{table:solids}

\newcolumntype{G}[1]{>{\raggedleft\arraybackslash}m{#1}}
\begin{threeparttable}
   \centering
   \setlength\extrarowheight{-6pt}
   \begin{tabular}{*{7}{G{0.07\textwidth}}m{0.01\textwidth}*{6}{G{0.07\textwidth}}}
        \tdrule
        
& & \multicolumn{5}{c}{PBE} & & \multicolumn{5}{c}{PBE0} \\

\cline{3-7}
\cline{9-13}\\[-3pt]
Solid & \multicolumn{1}{c}{Transition} & \texttt{stdlo} & \texttt{helo} & \multicolumn{1}{c}{Wien2K\tnote{1}} & FLEUR\tnote{2} & VASP\tnote{3} & & \texttt{stdlo} & \texttt{helo} & Wien2K\tnote{1} & FLEUR\tnote{2} & VASP\tnote{3} \\[5pt]

\midrule

Ar & $\Gamma \rightarrow \Gamma$ & 8.699&8.699 & 8.69 & 8.71 & 8.68& & 11.121 & 11.125 & 11.09 & 11.15 & 11.09\\
C & $\Gamma \rightarrow \Gamma$ & 5.599&5.599 & 5.59 & 5.64 & 5.59& & 7.702 & 7.701 & 7.69 & 7.74 & 7.69\\
 & $\Gamma \rightarrow \textrm{X}$ & 4.767&4.767 & 4.76 & 4.79 & 4.76& & 6.665 & 6.664 & 6.64 & 6.69 & 6.66\\
 & $\Gamma \rightarrow \textrm{L}$ & 8.467&8.467 & 8.46 & 8.58 & 8.46& & 10.776 & 10.776 & 10.76 & 10.88 & 10.77\\
Si & $\Gamma \rightarrow \Gamma$ & 2.562&2.562 & 2.56 & 2.56 & 2.57& & 3.955 & 3.955 & 3.95 & 3.96 & 3.97\\
 & $\Gamma \rightarrow \textrm{X}$ & 0.711&0.711 & 0.71 & 0.71 & 0.71& & 1.913 & 1.914 & 1.91 & 1.93 & 1.93\\
 & $\Gamma \rightarrow \textrm{L}$ & 1.537&1.537 & 1.53 & 1.54 & 1.54& & 2.861 & 2.861 & 2.86 & 2.87 & 2.88\\
GaAs & $\Gamma \rightarrow \Gamma$ & 0.543&0.543 & 0.53 & 0.55 & 0.56& & 1.985 & 1.994 & 1.99 & 2.02 & 2.01\\
 & $\Gamma \rightarrow \textrm{X}$ & 1.467&1.467 & 1.46 & 1.47 & 1.46& & 2.671 & 2.672 & 2.66 & 2.69 & 2.67\\
 & $\Gamma \rightarrow \textrm{L}$ & 1.015&1.014 & 1.01 & 1.02 & 1.02& & 2.358 & 2.363 & 2.35 & 2.38 & 2.37\\
MgO & $\Gamma \rightarrow \Gamma$ & 4.773&4.774 & 4.79 & 4.84 & 4.75& & 7.251 & 7.253 & 7.23 & 7.31 & 7.24\\
 & $\Gamma \rightarrow \textrm{X}$ & 9.140&9.141 & 9.16 & 9.15 & 9.15& & 11.617 & 11.620 & 11.58 & 11.63 & 11.67\\
 & $\Gamma \rightarrow \textrm{L}$ & 7.931&7.931 & 7.95 & 8.01 & 7.91& & 10.468 & 10.470 & 10.43 & 10.51 & 10.38\\
NaCl & $\Gamma \rightarrow \Gamma$ & 5.203&5.199 & 5.22 & 5.08 & 5.20& & 7.304 & 7.302 & 7.29 & 7.13 & 7.26\\
 & $\Gamma \rightarrow \textrm{X}$ & 7.584&7.582 & 7.59 & 7.39 & 7.60& & 9.820 & 9.821 & 9.80 & 9.59 & 9.66\\
 & $\Gamma \rightarrow \textrm{L}$ & 7.303&7.300 & 7.33 & 7.29 & 7.32& & 9.399 & 9.399 & 9.40 & 9.33 & 9.41\\

        \dtrule
    \end{tabular}
    \begin{tablenotes}
        \item [1] Reference \cite{Tran2011}
        \item [2] Reference \cite{Betzinger2010}
        \item [3] Reference \cite{Paier2006}
    \end{tablenotes}
\end{threeparttable}
  
\end{table*}

The results are shown in Tab.~\ref{table:solids} along with the data previously obtained with Wien2K, FLEUR and VASP. 
We find essentially perfect agreement between our PBE calculation performed with the \texttt{stdlo} and \texttt{helo} settings. The largest differences (2--4 meV) are obtained for the $\Gamma \rightarrow \Gamma$ transition in NaCl. 
We therefore conclude that our \texttt{stdlo} set is well-converged for PBE calculations.

The results from the present work agree to those of Wien2K up to $\sim\SI{10}{meV}$ for all solids, except for MgO and NaCl where the discrepancy is $\SIrange{20}{30}{meV}$. 
The mean absolute deviation (MAD) between our and Wien2K data is $\SI{11}{meV}$.
The difference between our and FLEUR data is much larger with the MAD of 43~meV exceeding 100~meV in several cases.
Such a disagreement goes above the precision thresholds defined in Sec.~\ref{sec:intro}.  


In PBE0 calculations, we find that the $\texttt{stdlo}$ and $\texttt{helo}$ settings still yield close results.
Only in the case of GaAs, the differences approach 10~meV for the $\Gamma\rightarrow\Gamma$ gap. 
Although the basis incompleteness in $\texttt{stdlo}$ barely influences these data, it is premature to generalize this result beyond the considered systems.  
On the other hand, we deduce that the absence of $\texttt{helo}$'s in the Wien2K and FLEUR calculations does not introduce a significant error.
The MAD between the Wien2K and our \texttt{helo} data is 16~meV, i.e., just slightly above that in the PBE calculations. 
The MAD with the FLEUR results is 54~meV, and, again, it is only slightly larger than in the PBE case. 
Thus, the main source of discrepancies appears already the PBE level suggesting that the Fock exchange calculation yield similar results despite large differences in the implementations.

\begin{figure}
\includegraphics[width=7.9cm]{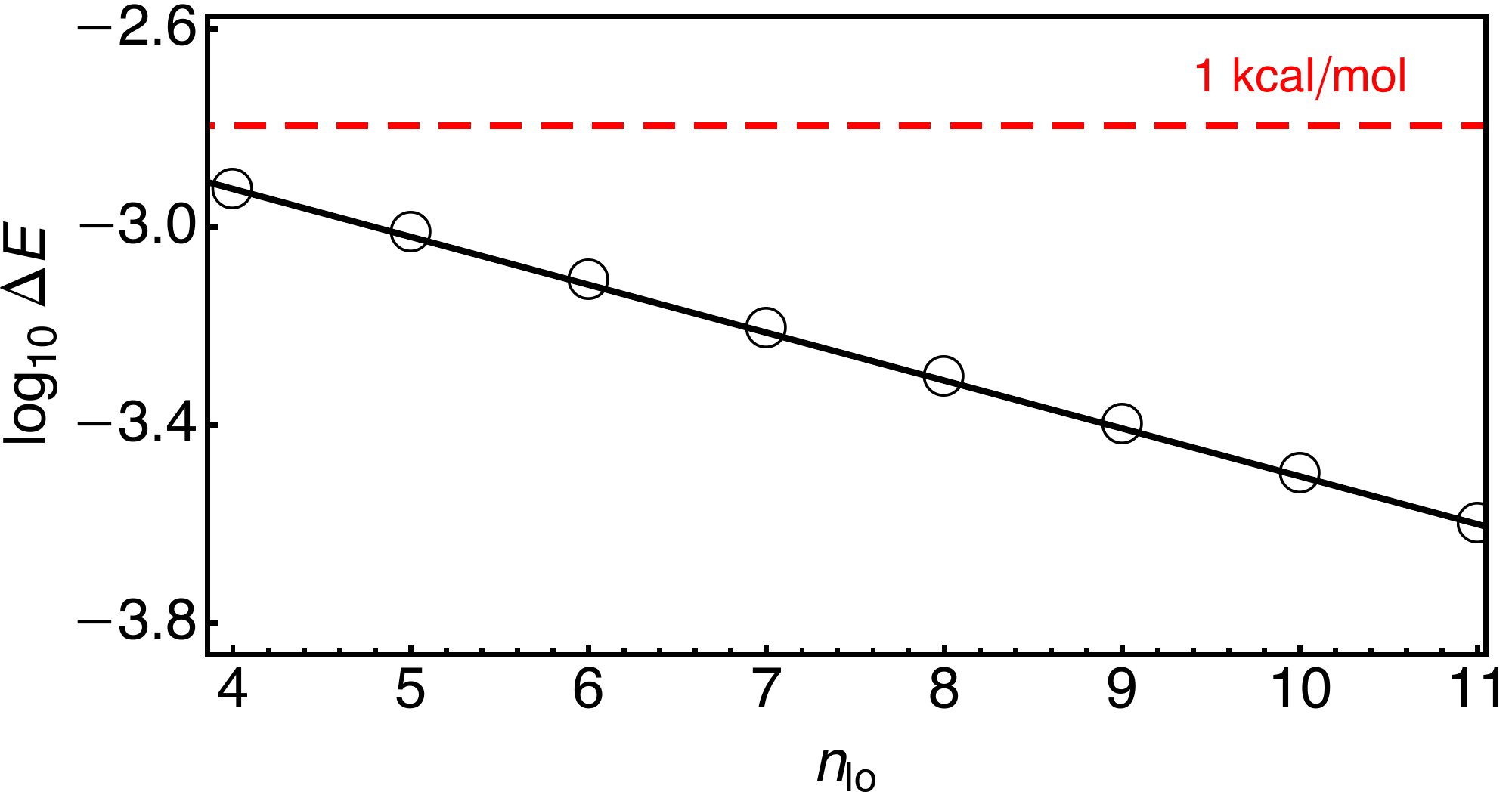}
\caption{\label{fig:Si_extranlo} 
Error in the PBE0 total energy of bulk silicon with respect to the number of additional local orbitals in the $l=0$ channel. The reference for the error calculation is the extrapolated value.}
\end{figure}

Although the additional LOs have a small impact on gaps, they still contribute to the total energy as in the Be calculation discussed in Sec.~\ref{sec:atoms}. 
Fig.~\ref{fig:Si_extranlo} shows how the error in the total energy reduces as \texttt{helo}'s are added to the $l=0$ channel. 
Similarly to the Be case, we observe an exponential decay although on a much larger scale. 

Finally, comparing our results with the VASP data obtained with the projector-augmented-waves method, we find the MAD of 11~meV and 31~meV in PBE and PBE0 calculations, respectively.
The maximum discrepancy of 160~meV was obtained in the PBE0 calculation for the $\Gamma \rightarrow X$ energy in NaCl, whereas it is only 18~meV in the PBE case. 
This result underlines the issue of having an inconsistency between the functional used in a simulation and generating a pseudopotential.

The previous implementation of hybrid functionals in \texttt{exciting} employs the BFB approach and relies on the mixed-basis formalism. 
Comparing performance of the present and previous implementations for the solids with the same number of unoccupied bands, we find similar job execution times on average despite applying just initial optimisations to our ACE code.
Techniques such as the interpolative separable density fitting have been shown to accelerate the ACE method dramatically in plane-wave calculations~\cite{Hu2017,Wu2022}.


\section{Conclusions}

In summary, we implemented the adaptively compressed exchange (ACE) operator within the LAPW framework.
ACE does not lead to any loss of precision even if the Fock exchange is applied to valence bands only, and we showed that this property applies in all-electron calculations too.
Therefore the ACE method compares favorably with previous hybrids implementations.
Our discussion was restricted to the Fock exchange, but the implementation can be extended to the screened and range-separated types of the exchange.

Our calculations reproduced non-relativistic PBE0 energies of atoms and molecules within $4~\mu$Ha/atom of the multi-resolution-analysis data.
To achieve this level of precision, it was important to expand the LAPW+LO basis with high-energy local orbitals, since the radial functions required for constructing the basis were obtained assuming a local PBE potential.
Thus, we proved that the benchmark qualities of LAPW extend beyond calculations with (semi)local exchange-correlation functionals.

Further, we applied the ACE code in a calculation of PBE0 gaps for 6 solids and found that high-energy local orbitals introduce a minor correction to the energies of valence-conduction transitions.
However, substantial corrections cannot be ruled out for materials beyond those considered here.
In comparison to other all-electron codes, we observed the differences with the Wien2K results that are below the targeted threshold of 10--20~meV for the most considered valence-conduction transitions. 

\section{Acknowledgments}
This work was funded by the Latvian Research Council via the project Precise Fock Exchange (PREFEX) with the grant agreement No. lzp-2020/2-0251.
D. Z. and A.G. also acknowledge the support from the European Union’s Horizon 2020 research and innovation program under Grant Agreement No. 951786 (NOMAD CoE).
The calculations were carried out in the HPC centre of Riga Technical University and LUMI supercomputer at CSC – IT center for science ltd. (pilot project No. 465000039).
Finally, we thank Prof. Stefan Goedecker for useful discussions and suggesting the wavelet-based Poisson solver.

%

\end{document}